\documentclass[12pt]{iopart}

\expandafter\let\csname equation*\endcsname\relax
\expandafter\let\csname endequation*\endcsname\relax
\usepackage{amsmath}
\usepackage{bm}

\usepackage{graphicx,color}

\usepackage{caption}
\usepackage{subcaption}

\usepackage{tikz}
\usetikzlibrary{quantikz}

\begin{document}

\title[Global optimization in variational quantum algorithms]{Global optimization in variational quantum algorithms via dynamic tunneling method}

\author{Seung Park$^1$, Kyunghyun Baek$^2$, Seungjin Lee$^{*2}$, Mahn-Soo Choi$^{\dagger 1}$}
\address{$^1$Department of Physics, Korea University, Seoul 02841, Republic of Korea}
\address{$^2$Electronics and Telecommunications Research Institute, Daejeon 34129, Republic of Korea}

\eads{\mailto{$^*$seungjin.lee@etri.re.kr}, \mailto{$^\dagger$choims@korea.ac.kr}}

\begin{abstract}

 We present a global optimization routine for the variational quantum algorithms, which utilizes the dynamic tunneling flow. Originally designed to leverage information gathered by a gradient-based optimizer around local minima, we adapt the conventional dynamic tunneling flow to exploit the distance measure of quantum states, resolving issues of extrinsic degeneracy arising from the parametrization of quantum states. Our global optimization algorithm is applied to the variational quantum eigensolver for the transverse-field Ising model to demonstrate the performance of our routine while comparing it with the conventional dynamic tunneling method, which is based on the Euclidean distance measure on the parameter space.

\end{abstract}


\submitto{\NJP}

\maketitle

\section{Introduction}
\label{sec:intro}
The variational quantum algorithm (VQA) is a hybrid quantum-classical algorithm that has attracted significant attention in recent years owing to its potential to address complex problems in quantum chemistry \cite{peruzzo2014variational, McArdle2020chemistry}, material science \cite{Ma2020material, Endo2020material}, physics \cite{Bauer2023physics}, computational science \cite{jager2023universal, amaro2022case}, and many other areas. 
By combining classical optimization techniques with quantum computation, the VQA leverages the variational principle of quantum mechanics to find the minimum of the cost function \cite{Bharti2022review, Cerezo2021review}.

In mathematical terms, the VQA can be framed as an optimization problem, where the cost function is defined as the expectation value of an observable with respect to a variational quantum state, referred to as an ansatz state, implemented through a parametric quantum circuit.
Classical optimization techniques are then applied to iteratively update the parameters of the parametric quantum circuit until the expectation value of the observable converges to the lowest value.

As an optimization problem, the choice of classical optimizer plays a critical role in efficiently solving the VQA, along with the design of a suitable ansatz state. While well-established classical optimizers are directly utilized, several optimizers have been developed specifically to enhance VQA's performance, focusing on gradient-based optimization routines due to their proven convergence properties and the ability of quantum circuits to implement the gradient of VQA in parallel. Among the optimizers explored are the stochastic gradient descent \cite{kingma2017adam, sweke2020}, the Riemannian gradient flow \cite{Wiersema2023Riemannian}, and the natural gradient method \cite{Stokes2020natural, Wierichs2020}.

However, gradient-based optimizers for VQA often encounter the challenge of local minima, which is inherent both in gradient-based optimization and the variational method of quantum mechanics.
Various designs for ansatz have been proposed to mitigate such local convergence, notably including \cite{PhysRevResearch.3.023092,10.1116/5.0162455,Park2024hamiltonian,10.1063/5.0186205}.
Among other things, it has been suggested to over-parameterize quantum state \cite{Larocca2023overparameterization}, introducing more quantum layers to transform local minima into saddle points, akin to strategies used in classical neural networks \cite{Livni2014, Li2018}. However, over-parameterization presents practical challenges, including the limitations imposed by noise in NISQ devices and their restricted coherence time.
Moreover, increasing circuit depth can exacerbate local minima issues, expanding the search space \cite{Sim2019expressibility, Holmes2022expressibility} and potentially leading to barren plateaus \cite{Wierichs2020, cerezo2021cost}. 

Consequently, global optimization routines have been proposed to complement the design of parametric quantum states, with most methods being gradient-free to avoid the drawbacks of gradient-based optimization. Techniques like the Nelder-Mead method \cite{nelder1965simplex, guerreschi2017practical}, bound optimization by quadratic approximation \cite{powell2009bobyqa, lavrijsen2020classical}, and the quantum kernel surrogate model-based method \cite{smith2023faster} have been explored, though they may require a significant number of function evaluations to converge, raising scalability concerns in the VQA context.

In this work, we propose a global optimization strategy specialized for VQA that incorporates a gradient-based optimizer called the dynamic tunneling method \cite{Yao1989}. Similar to the conventional dynamic tunneling method, the optimizer uses the local minima detected by the optimizer to generate a dynamic flow towards a global minimum. 
However, 
our modified dynamic tunneling method enhances efficiency by addressing the limitations of the conventional method within the VQA framework.

This paper is organized as follows: In the next section, we first introduce the conventional dynamic tunneling method and then propose its modification for VQA, incorporating a distance measure between quantum states. In \Sref{sec:examples}, we demonstrate the global optimization of the transverse field Ising model using the modified dynamic tunneling method, comparing its performance with the conventional dynamic tunneling method applied to the same problem. \Sref{sec:outlook} is devoted to the conclusion and outlook of our work.

\section{Dynamic tunneling method on VQA}
\label{sec:dtvqe}

\subsection{Dynamic tunneling method}
\label{sec:general_dtopt}

Dynamic tunneling method and its variants \cite{Yao1989, cetin1993} are global optimization algorithms implementing a dynamic flow to escape from a valley around a stable point to a valley around another stable point having the lower value of the cost function of an optimization problem.
In its primitive form, this is achieved by constructing a so-called energy function from the cost function by exploiting the information of local minima of the cost function.

Explicitly, given an cost function $f(\bm{x})$, the dynamic tunneling flow is constructed by promoting optimization parameters $\bm{x}$ to be a flow $\bm{x}(t)$ generated by
\begin{align} \label{eq:tunneling_flow}
 \dot{\bm{x}} = - \frac{\partial E}{\partial \bm{x}}
\end{align}
where $E = E(\bm{x};\bar{\bm{x}}, \lambda, k)$ is an enegy function given by \cite{Yao1989}
\begin{align} \label{eq:tunnel_energy}
 E(\bm{x};\bar{\bm{x}}, \lambda, k) = \frac{f(\bm{x})-f(\bar{\bm{x}})}{|\bm{x}-\bar{\bm{x}}|^{2\lambda}}+k\int_0^{f(\bm{x})-f(\bar{\bm{x}})}F_{\textrm{ReLU}}(z)dz.
\end{align}
In \eref{eq:tunnel_energy}, $\lambda$ and $k$ are hyperparameters depending on the problem, and $\bar{\bm{x}}$ is a stable point (or a local minimum) of the cost function. Also, $F_{\textrm{ReLU}}(z)$ is the rectified linear unit which is zero only when the $z \le 0$.
Together with an appropriately chosen $\lambda$, the first term in \eref{eq:tunnel_energy} amounts to induce a pole of the energy function at $\bar{\bm{x}}$ while violating the Lipschitz condition at $\bar{\bm{x}}$, which enables the flow to converge to a point outside of the valley around $\bar{\bm{x}}$ \cite{Yao1989,cetin1993}. On the other hand, the second term in \eref{eq:tunnel_energy} is a penalty term with weight $k$ imposing the tunneling to seek a point having a lower cost value than $f(\bar{\bm{x}})$.
Instead of constructing an energy function for a tunneling flow, one can introduce a dynamic flow directly via \cite{Yao1989}
\begin{equation} \label{eq:tunnel_dynamic}
\dot{\bm{x}}=-\left(\frac{1}{|\bm{x}-\bar{\bm{x}}|^{2\lambda}}+k\,F_{\textrm{ReLU}}(f(\bm{x})-f(\bar{\bm{x}}))\right)\frac{\partial f(\bm{x})}{\partial \bm{x}}
\end{equation}
which is more convenient for a practical implementation. Clearly, the dynamic flow in \eref{eq:tunnel_dynamic} has the same pole structure and penalty constraint as the energy function in \eref{eq:tunnel_energy}.

Having a dynamic tunneling flow, the conventional dynamic tunneling method corresponds to a gradient-based optimization algorithm followed by an implementation of the dynamic tunneling flow in \eref{eq:tunnel_dynamic} for the stable point recognized by the optimization process. In turn, the dynamic tunneling method is a successive application of the optimization-tunneling pair until the tunneling flow can find no lower valley.

\begin{figure}
     \centering

     \begin{subfigure}{0.49\textwidth} 
         \centering
         \includegraphics[width=7.5cm]{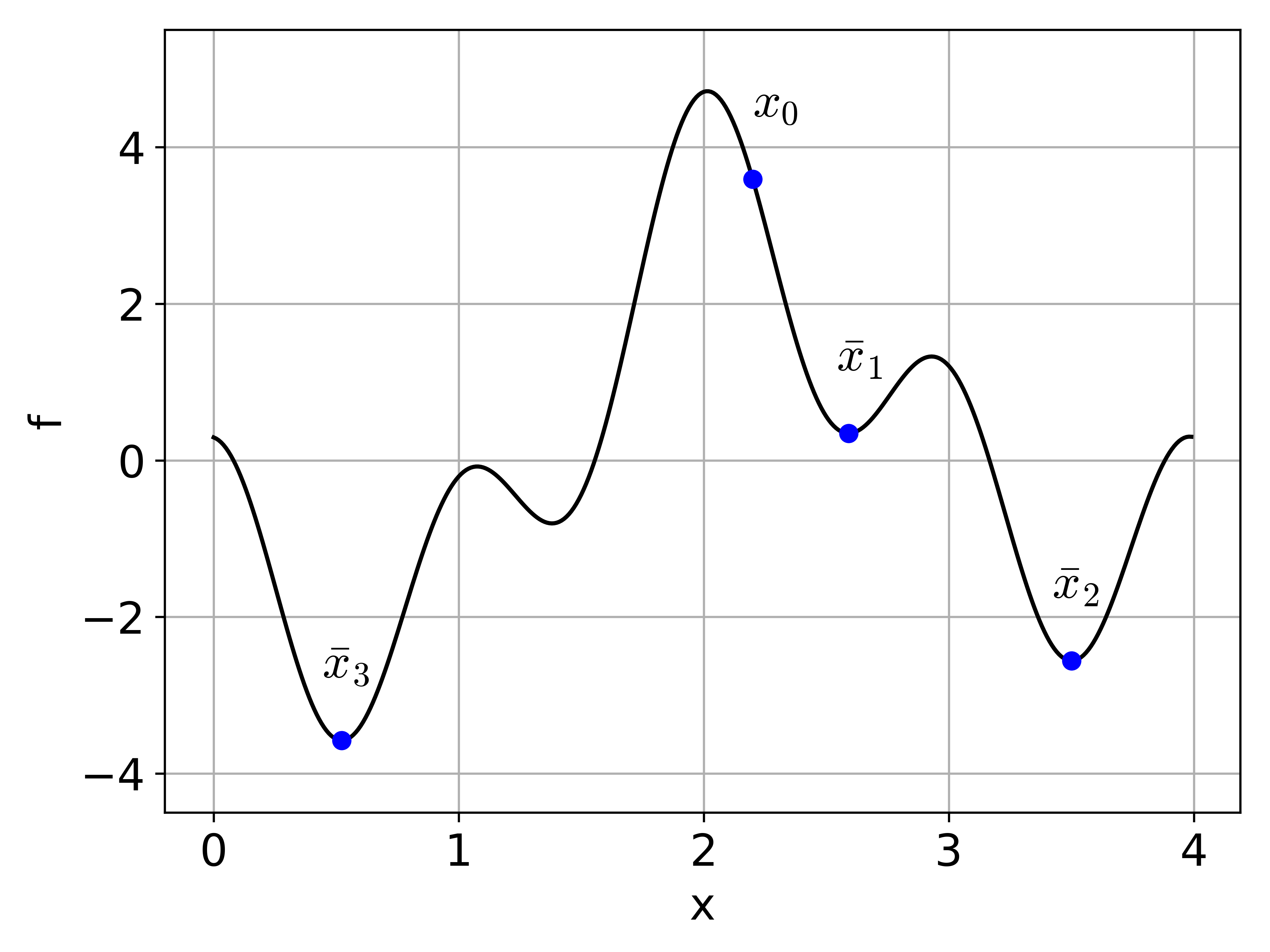}
        \caption{}
       
     \end{subfigure}
     \hfill
     \begin{subfigure}{0.49\textwidth} 
         \centering
         \includegraphics[width=7.5cm]{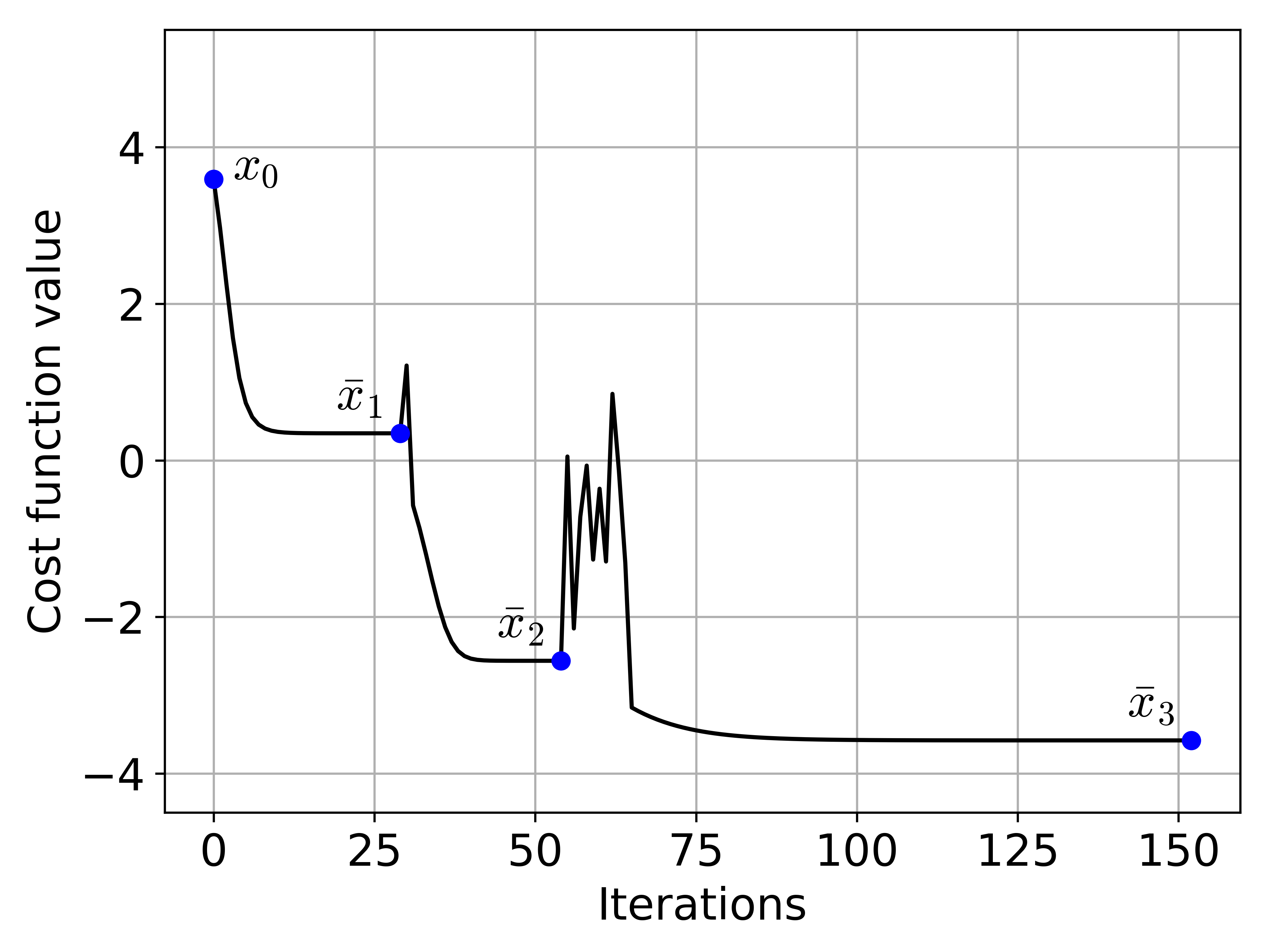}
       \caption{}
       
     \end{subfigure}
\caption{An illustration of the dynamic tunneling flow. (a) The landscape of a cost function. (b) The change of the cost function as iteration proceeds. The blue dots in (b) represent the corresponding points on the landscape in (a). The sharp peaks in (b) indicate the occurrence of dynamic tunneling.}
\label{example}
\end{figure}
\textbf{}

To illustrate the dynamic tunneling method and point out our method more clearly below,
let us consider an optimization problem defined by the cost function
\begin{equation}\label{eq:f_example}
 f(x)=\cos\left(\frac{\pi}{2}\left(x-\frac{1}{2}\right)\right)-\frac{1}{2}\cos\left(2\pi\left(x+\frac{3}{2}\right)\right)+\sin\left(\pi\left(x+\frac{1}{2}\right)\right)-\frac{3}{2}\sin\left(\frac{\pi}{2}(x+1)\right),
\end{equation}
which is periodic in $x\in [0,4)$. The landscape of \eref{eq:f_example} is plotted in \Fref{example}(a).
We first employ the usual gradient-descent optimizer with step size $0.005$ and the initial parameter $x_{0} = 2.2$ indicated as the point $x_{0}$ in \Fref{example}(a). The optimizer then ends up with a stable point $\bar{x}_{1}$ in \Fref{example}(a). Given a local minimum $\bar{x}_{1}$, which now becomes the initial point for the next optimization process, a dynamic tunneling flow \eref{eq:tunnel_dynamic} sets in
and the optimizer escapes from $\bar{x}_{1}$,
leading to the next stable point $\bar{x}_{2}$ depicted in \Fref{example}(a).
By repeatedly alternating the relaxation to a local stable point by gradient-descent and the subsequent escape from it by dynamic tunneling, the whole dynamic tunneling routine traverses from the initial point $x_0$ through local stable points $\bar{x}_{1}$, $\bar{x}_{2}$, and $\bar{x}_{3}$, among which $\bar{x}_{3}$ is the global minimum of the problem, as illustrated in \Fref{example}(b). 
For the hyperparameters, in this particular example, we have chosen $k=50$ and set $\lambda$ to be $1.5$ times the lowest power of gradient of convergence to each stable point.
The lowest power is extracted from the last two iterations reaching each local minimum.

\subsection{Dynamic tunneling method with distance measure for quantum states}
\label{sec:dtopt_vqe}

The VQA is a quantum algorithm that implements the variational principle of quantum mechanics by combining a parametric quantum circuit and a classical optimization algorithm \cite{peruzzo2014variational, mcclean2016theory, cerezo2021variational}. In its simplest form, a VQA can be cast into the form of
\begin{align} \label{eq:quantum_function}
    f(\bm{x}; O) = \textrm{Tr}\,(O \rho(\bm{x}))
\end{align}
where $O$ and $\rho$ are a target observable and the density operator of a quantum state depending on a parameter $\bm{x}$, respectively. In practice, particularly for an $N$-qubit system, the quantum state in \eref{eq:quantum_function} is prepared by applying the parametrized quantum circuit implementing a unitary operator $U(\bm{x})$ to a fixed initial state $\rho_{0}$ so that the parametric quantum state has the form of
\begin{align}
    \rho(\bm{x}) = U(\bm{x}) \rho_{0} U(\bm{x})^{\dagger},
\end{align}
often called an ansatz state.

The quantum function $f(\bm{x}; O)$ corresponds to a parametrized quantum expectation on the observable $O$, so an implementation of the variational method can be obtained by optimizing $f(\bm{x};O)$ for $\bm{x}$ primarily by employing a classical optimizer. Consequently, the VQA is an optimization problem whose cost function corresponds to the quantum function of a parametric state, 

By identifying a VQA as an optimization problem defined by a quantum function, we can apply the dynamic tunneling method in \Sref{sec:general_dtopt} by introducing a tunneling flow as in \eref{eq:tunneling_flow}. However, apart from the intrinsic degeneracy of the target observable, the quantum function is not extrinsically injective since multiple parameters can induce the same quantum state due to the periodic feature of the unitary gates. In turn, the dynamic flow starting from a local minimum, say $\bar{\bm{x}}$, may converge to another point $\bar{\bm{x}}'$ with $\rho(\bar{\bm{x}}')=\rho(\bar{\bm{x}})$ which makes the algorithm computationally redundant.

Such extrinsic degeneracy thus not only makes the algorithm computationally inefficient but also prevents the termination criterion of the algorithm from working properly since we cannot exploit the divergence of the tunneling flow to ensure the global minimum. Therefore, instead of using the parameter-space distance $|\bm{x}-\bar{\bm{x}}|$, we employ a distance measure on the space of quantum states in the first term of \eref{eq:tunneling_flow}. Explicitly, the tunneling flow for VQA is given by
\begin{align} \label{eq:tunneling_flow_vqe}
\dot{\bm{x}}=-\left(\frac{1}{\mathcal{D}(\rho(\bm{x}), \rho(\bar{\bm{x}}))^{2\lambda}}+k\,F_{\textrm{ReLU}}(f(\bm{x})-f(\bar{\bm{x}}))\right)\frac{\partial f}{\partial \bm{x}}
\end{align}
where $\mathcal{D}(\rho,\rho')$ is a distance measure on states $\rho$ and $\rho'$. 

By introducing the distance measure on quantum states, one can avoid all other local minima corresponding to the same quantum state since the modified dynamic tunneling flow completely excludes the quantum state itself due to the first term inside the parenthesis in \eref{eq:tunneling_flow_vqe}.
It is emphasized that the exclusion globally occurs over the whole parameter space,
enabling the flow to circumvent a whole set of local minima corresponding to each quantum state.

There are several distance measures for quantum states \cite{choi2022quantum}, and most of them have a similar effect when they are cast into the dynamic tunneling method. Here, we utilize the Hilbert-Schmidt (HS) distance as the distance measure for the dynamic tunneling method. The Hilbert-Schmidt distance $\mathcal{D}_{\textrm{HS}}$ between two density matrices $\rho$, $\sigma$ is given by 
\begin{equation}
\mathcal{D}_{\textrm{HS}}(\rho,\sigma)= \sqrt{\textrm{Tr}\left((\rho-\sigma)^\dagger(\rho-\sigma)\right)} .
\end{equation}
If the two density matrices are of pure states, denoted by $\rho\equiv|\psi\rangle\langle\psi|$ and $\sigma\equiv|\phi\rangle\langle\phi|$, the Hilbert-Schmidt distance between them can be expressed as 
\begin{equation}
\mathcal{D}_{\textrm{HS}}(|\psi\rangle\langle\psi|,|\phi\rangle\langle\phi|)=\sqrt{2-2|\langle\psi|\phi\rangle|^2} .
\end{equation}
Consequently, in most cases of the VQA, the distance can be readily obtained by measuring the fidelity between the two states \cite{flammia2011}.

In addition to resolving the extrinsic degeneracy of the quantum function, the Hilbert-Schmidt distance is practically appreciable since it is bounded by $\sqrt{2}$, regardless of the size of the parameter space. In particular, due to the upper bound, the denominator of the first term in \eref{eq:tunneling_flow_vqe} does not significantly affect the gradient of the cost function, thereby ensuring the convergence of the tunneling process. 

Moreover, local minima with different values of quantum function correspond to nearly orthogonal quantum states, and hence one can expect that the distance at the end of the tunneling process is close to the upper bound, $\sqrt{2}$. This facilitates the estimation of appropriate values for the hyperparameters of the optimization, such as the learning rate, in contrast to the tunneling algorithm using the parameter-space distance in which the distance between any two local minima is unpredictable.

\section{Case study: transverse-field Ising model}
\label{sec:examples}

We test the performance of the method with numerical simulation on the variational quantum eigensolver (VQE), one of the main applications of the VQA that seeks the ground-state energy of the given Hamiltonian. In this test, we simulate the VQE on the one-dimensional (1D) transverse-field Ising model. 

The transverse-field Ising model, consisting of spins arranged in a lattice, is a prototype model exhibiting various correlation effects in many-body physics and has been extensively studied in condensed matter physics \cite{pfeuty1970one, rieger1994zero, fisher1995critical}. Furthermore, the one-dimensional version allows for an exact solution \cite{pfeuty1970one}. This makes the model an excellent benchmark for many quantum algorithms, including the VQE.

The transverse-field Ising model is described by the Hamiltonian
\begin{equation}\label{eq:tfi_1d}
H=J\sum_{j=1}^{N-1} \sigma_j^{z}\sigma_{j+1}^z+B\sum_{j=1}^N\sigma_j^x ,
\end{equation}
where $N$ is the number of spins (i.e. qubits), $J>0$ is the coupling strength between the nearest-neighbor spins and $B$ is the strength of the external magnetic field in the transverse direction. We set $J=1$, therefore all the energy will be measured in units of $J$. 
In the tests below, we naturally choose the Hamiltonian as a target observable.

Note that the model exhibits a quantum phase transition from the ordered (i.e., anti-ferromagnetic) to disordered (i.e., paramagnetic) states. This phase transition may be understood in terms of spontaneous breaking of $Z_2$ symmetry:
To see this, note that the Hamiltonian is invariant under the symmetry transformation $X\equiv \prod_j \sigma_j^x$, that is, $[H,X]=0$. In the $B\to0$ limit (ordered phase), the ground states of the model are two-fold degenerate and are related to each other by the symmetry transformation $X$. On the other hand, in the $B\to\infty$ limit (disordered phase), the ground state is non-degenerate; the symmetry transformation $X$ maps the ground state to itself.
In this work, we focus on the disordered (paramagnetic) phase to avoid unnecessary complications due to the ground-state degeneracy.

\begin{figure}

\begin{center}
    \begin{quantikz}[row sep = 10]
        \lstick{\ket{0}} & \gate{R_{y}\left( \phi_{1} \right)} & \ctrl{1} & \qw      & \gate{R_{y}\left( \phi_{6} \right)} & \ctrl{1} & \qw      & \qw\\
        \lstick{\ket{0}} & \gate{R_{y}\left( \phi_{2} \right)} & \targ{ } & \ctrl{1} & \gate{R_{y}\left( \phi_{7} \right)} & \targ{ } & \ctrl{1} & \qw\\
        \lstick{\ket{0}} & \gate{R_{y}\left( \phi_{3} \right)} & \ctrl{1} & \targ{ } & \gate{R_{y}\left( \phi_{8} \right)} & \ctrl{1} & \targ{ } & \qw\\
        \lstick{\ket{0}} & \gate{R_{y}\left( \phi_{4} \right)} & \targ{ } & \ctrl{1} & \gate{R_{y}\left( \phi_{9} \right)} & \targ{ } & \ctrl{1} & \qw\\
        \lstick{\ket{0}} & \gate{R_{y}\left( \phi_{5} \right)} & \qw      & \targ{ } & \gate{R_{y}\left( \phi_{10} \right)}& \qw      & \targ{ } & \qw
    \end{quantikz}
\end{center}

\caption{A parameterized quantum circuit corresponding to a variational ansatz for 5 qubits. $R_y(\phi_i):=e^{-i \sigma^y \phi_i/2}$ denotes the single-qubit rotation around the $y$-axis.}
\label{ansatz}
\end{figure}
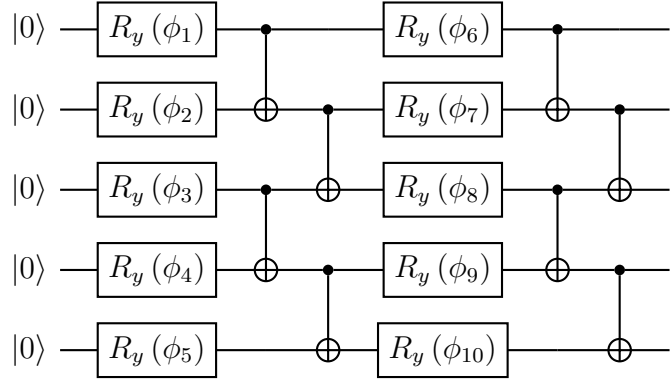

Here, we consider the case of $N=10$ and $B=5$, in which the system is in the nondegenerate, paramagnetic phase. The ground-state energy is about $-50.45$. 
Note that the Hamiltonian is real-valued in the computational basis. This implies that the wave function of any eigenstate of the Hamiltonian can be chosen to be real in the same basis.
We exploit this fact to reduce the variational ansatz subspace.
Figure~\ref{ansatz} shows the parameterized quantum circuit to generate our variational ansatz for $N=5$; for larger systems, the quantum circuit can be extended in a similar pattern. Each $R_y$ involves the trainable parameter $\phi_i$, while the structure of the CNOT gates mirrors the nearest-neighbor connectivity of the system, thereby creating entanglement.

\begin{figure}
     \centering

     \begin{subfigure}{0.49\textwidth} 
         \centering
         \includegraphics[width=7.5cm]{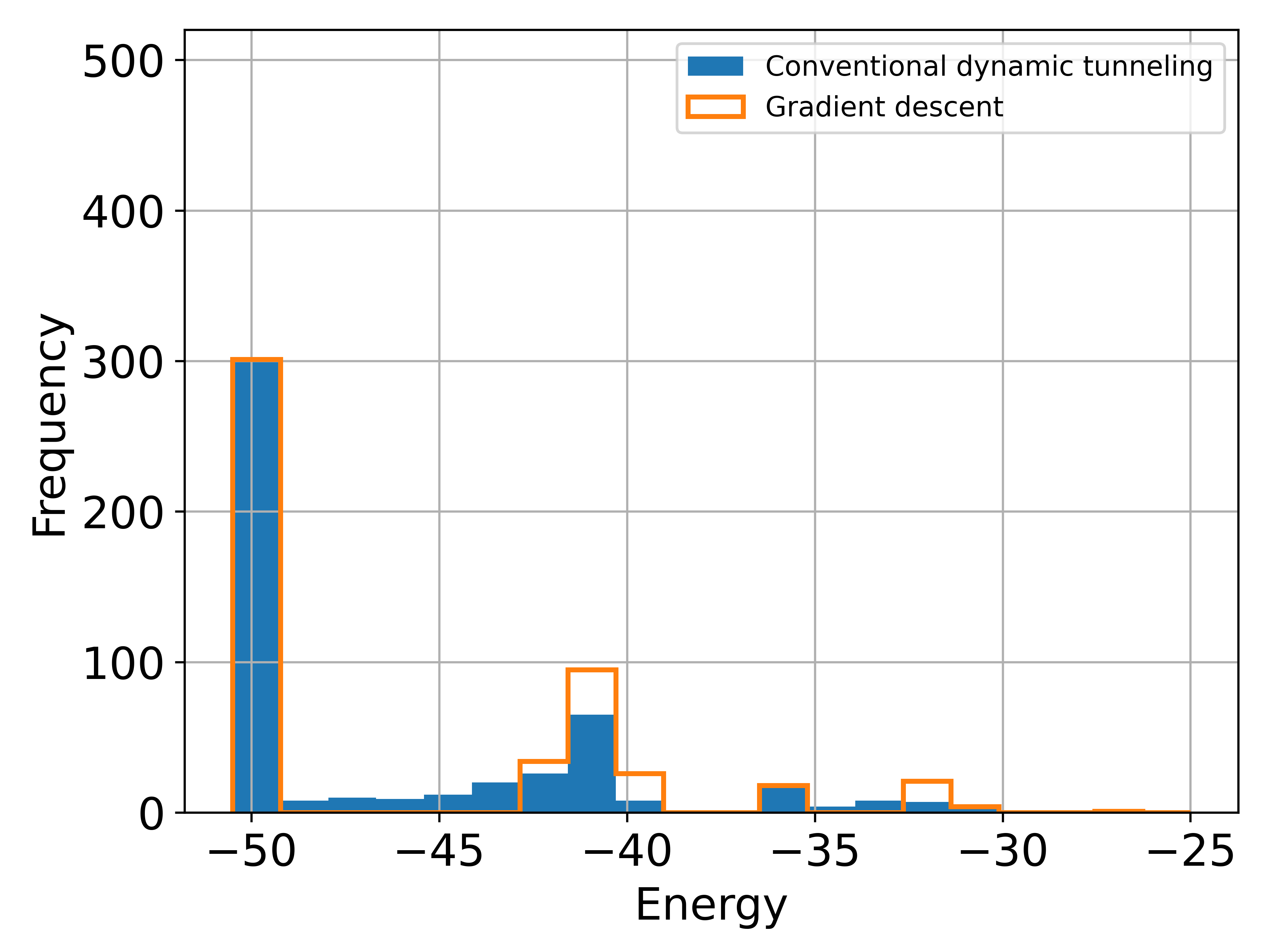}
        \caption{}
        
     \end{subfigure}
     \hfill
     \begin{subfigure}{0.49\textwidth} 
         \centering
         \includegraphics[width=7.5cm]{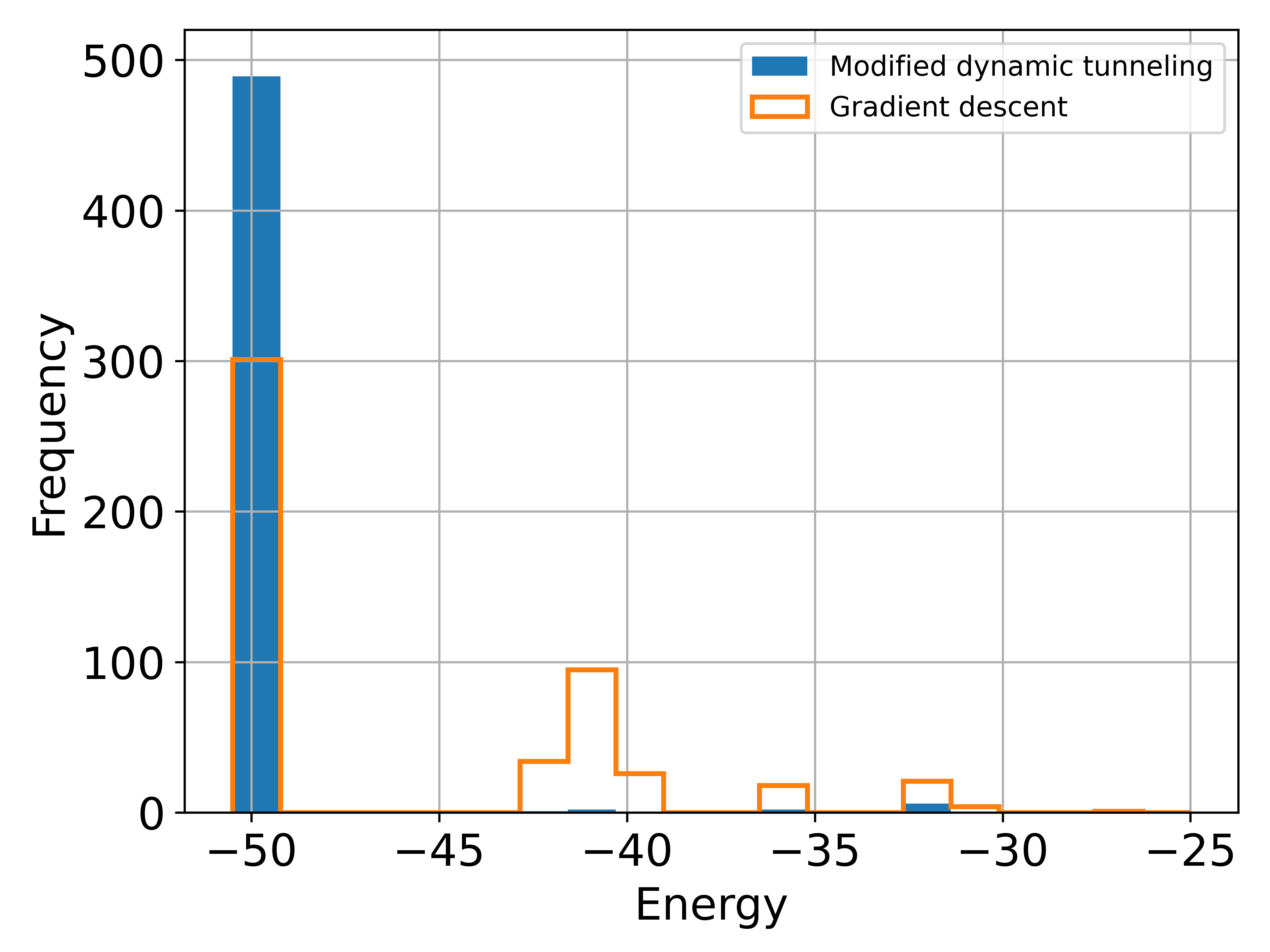}
       \caption{}
       
     \end{subfigure}
     
\caption{A comparison of the overall convergence of the (a) conventional and (b) modified dynamic tunneling methods. Each histogram shows the final converged values of the cost function over 500 random samples.
For a reference, we also show the histogram with empty bars and orange boundaries from the simple gradient-descent method. The global minimum is approximately -50.45 in this example.}
\label{hist}
\end{figure}

\Fref{hist} summarizes our main results, comparing the performances of the simple gradient-descent optimization algorithm (empty boxes with orange boundaries, in both (a) and (b)), (a) the conventional dynamic tunneling algorithm based on the parameter-space distance and (b) the modified dynamic tunneling algorithm based on the distance measure on quantum states.

We have set the hyperparameters for our simulation as follows: $\lambda$ for each tunneling is chosen to be $1.5$ times the lowest power of gradient of convergence to each stable point. $k=1.125$, and the learning rate is $0.01$, fixed over the whole process. For practical reasons, we limit the number of execution of the dynamic tunneling algorithm to 6 times, with the iteration limit of $2000$ for each run.

In \Fref{hist}(b), we can see that the modified dynamic tunneling method successfully escapes the local minima and reaches the global minimum, while the conventional one fails to converge into the global minimum as shown in \Fref{hist}(a). A few results that remained in local minima in \Fref{hist}(b) can also reach the global minimum by increasing the iteration limit.

\begin{figure}
     \centering
     \begin{subfigure}{0.49\textwidth} 
         \centering
         \includegraphics[width=7.5cm]{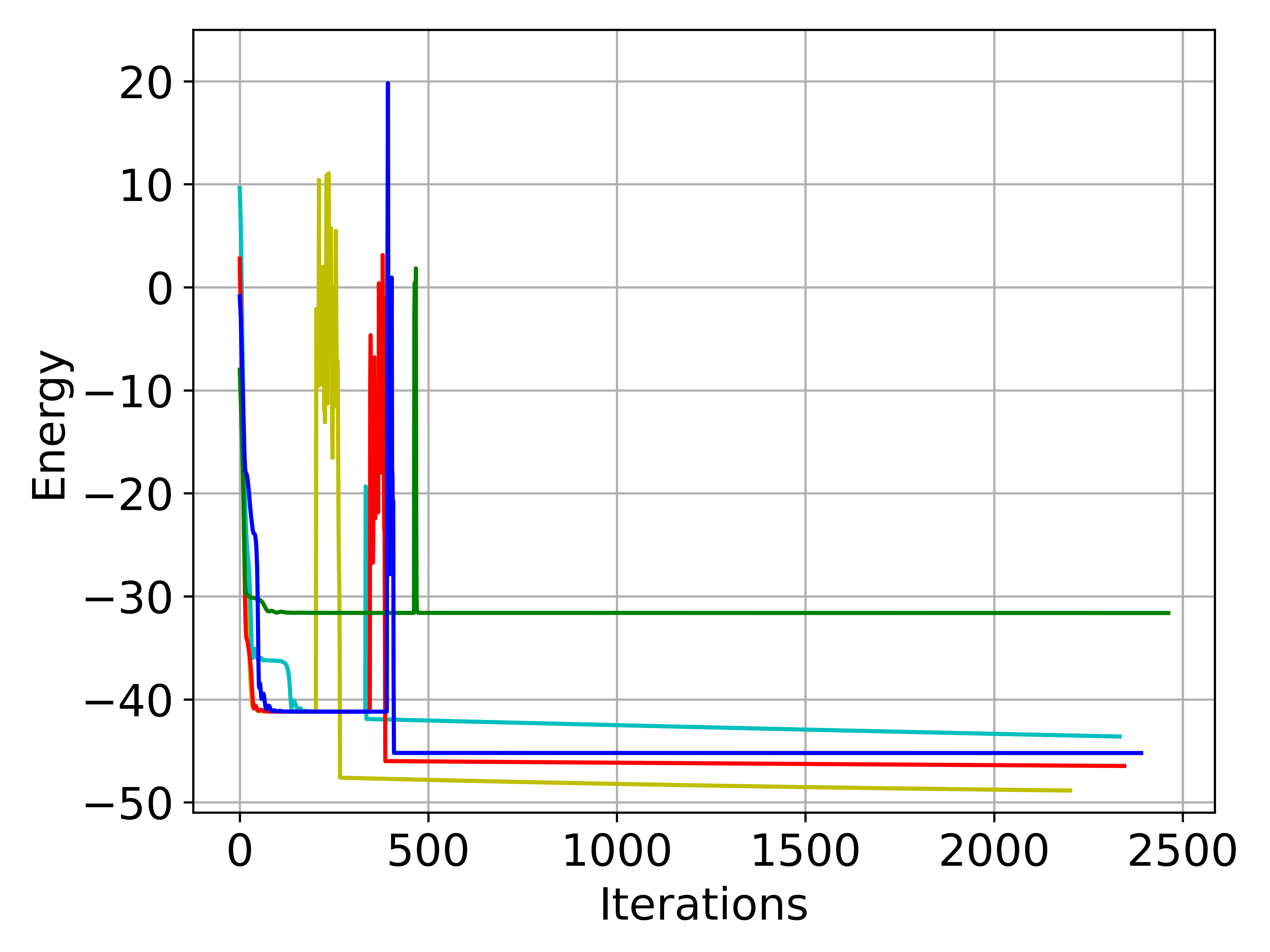}
        \caption{}
        
     \end{subfigure}
      \hfill
     \begin{subfigure}{0.49\textwidth}
         \centering
         \includegraphics[width=7.5cm]{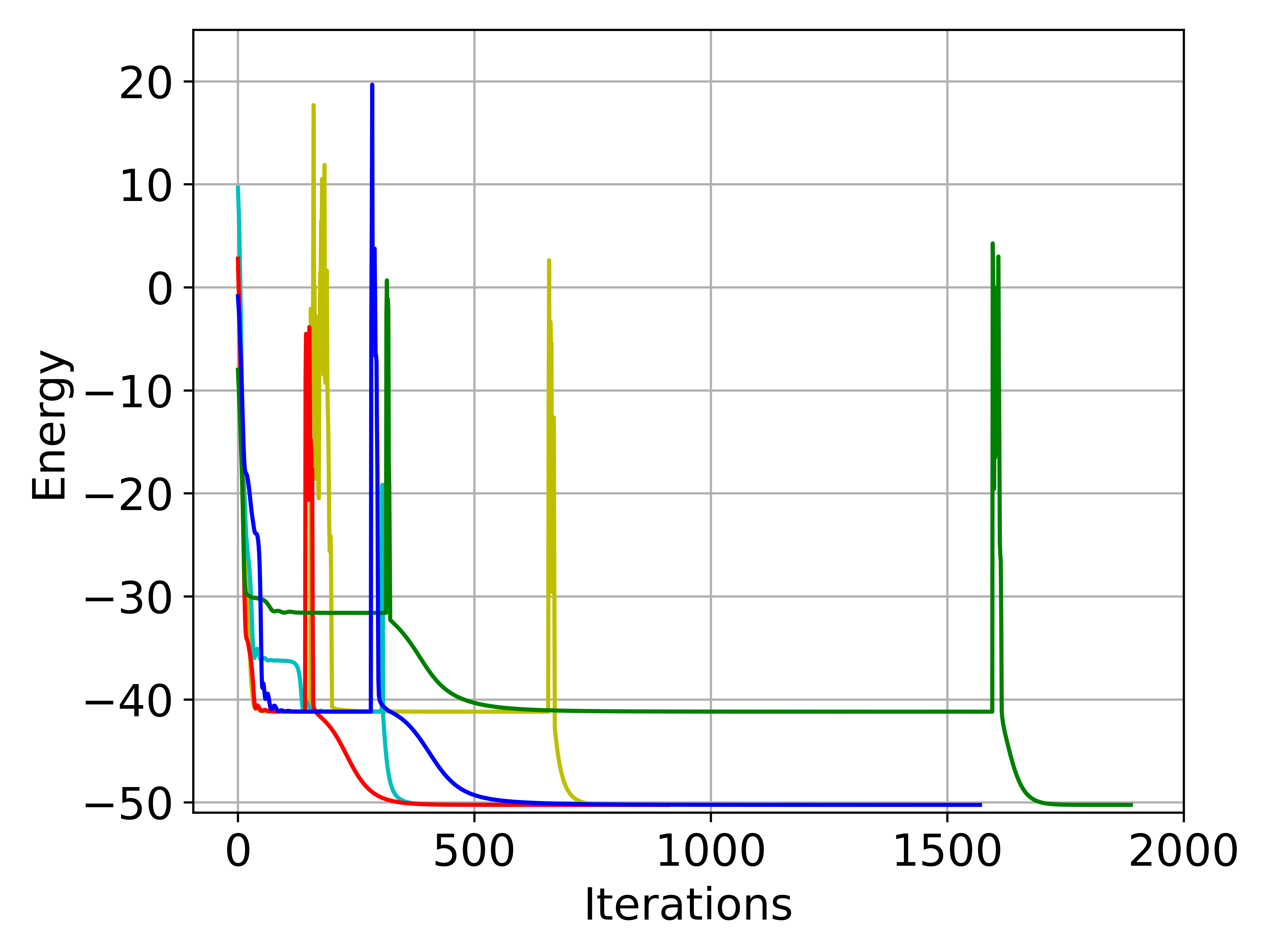}
        \caption{}
        
     \end{subfigure}
\caption{Iteration profiles of the cost function (energy) over the entire optimization process in the (a) conventional and (b) modified dynamic tunneling methods for 5 representative samples out of the whole in \Fref{hist}. Each sharp peak indicates the occurrence of dynamic tunneling.}
\label{convplot}
\end{figure}

\Fref{convplot} presents typical profiles of the cost function over the entire optimization process. It is interesting to note that in \Fref{convplot}(a), each optimization procedure stops even on a slope rather than at local minima on the landscape of the cost function in the parameter space, indicating that the tunneling process does not converge within the given iteration limit. This is the reason why the histograms in \Fref{hist}(a) are smoothly distributed over the global and local minima.
On the other hand, the iterations for the modified dynamic tunneling converge well first to local minima and eventually to the global minimum, as shown in \Fref{convplot}(b).

\begin{figure}
 \centering
 \includegraphics[width=12cm]{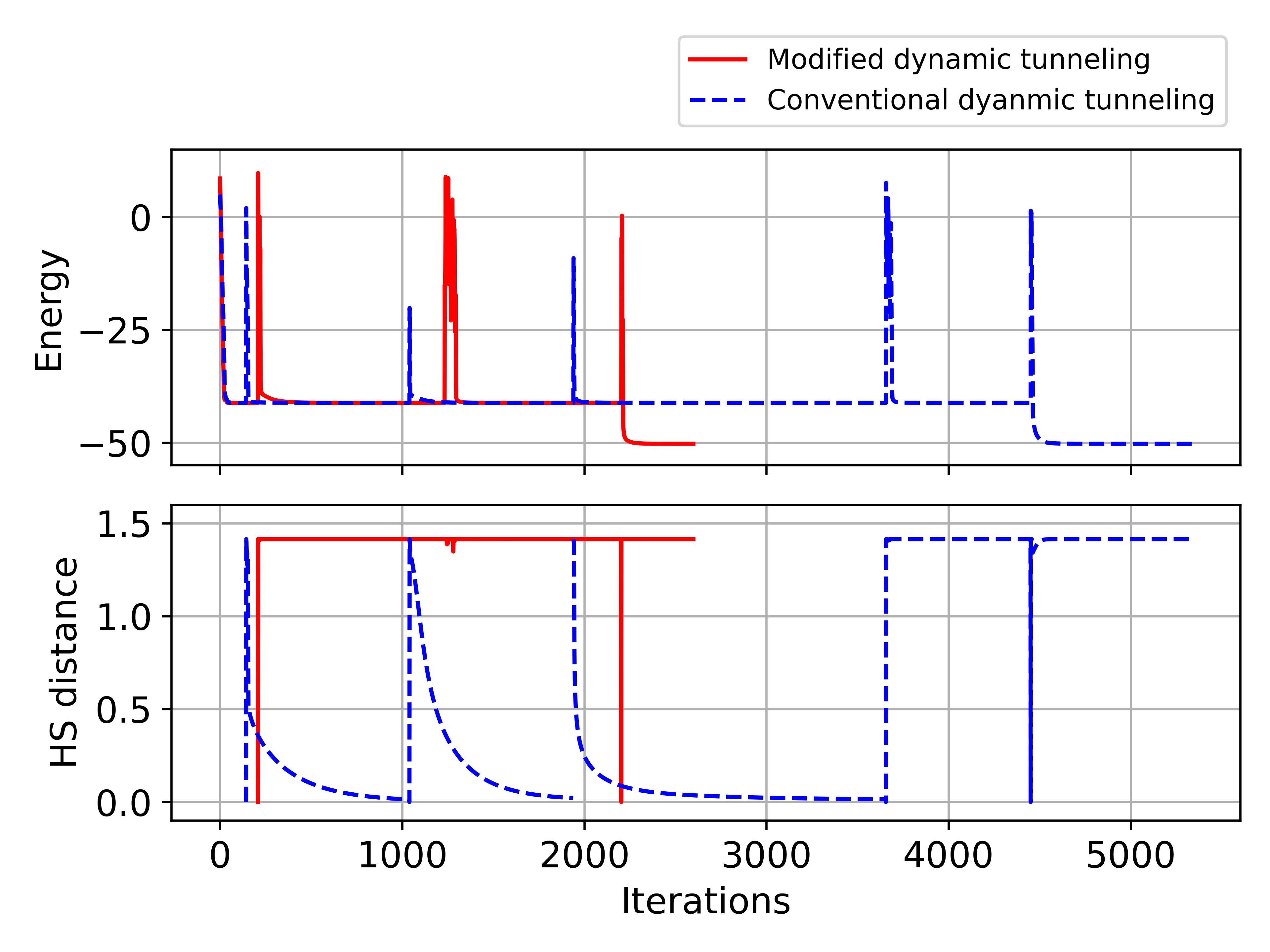}
\caption{A detailed comparison of the iteration profiles of the conventional (blue dashed curves) and modified (red curves) dynamic tunneling methods for selected worst-case samples. For the conventional method, we have put an ad-hoc bound $2^\lambda$ on the $\lambda$ term to avoid the convergence problem. Even with such a recipe, the conventional method converges much slower (almost twice as slow) than the modified method due to the extrinsic degeneracy.}
\label{lmsameqstates}
\end{figure}

Further details of the dynamical aspects of the conventional and modified dynamical tunneling algorithms are compared in \Fref{lmsameqstates}.
As one can see from the blue curves in \Fref{lmsameqstates}, the conventional dynamic tunneling method often encounters local minima of the same quantum state. In our simulation, this happened 57 times per 431 samples which start the tunneling process from local minima. \Fref{lmsameqstates} shows the worst-case scenario observed during the modified tunneling optimization over the 431 samples. One can see that even when the flow converges into the local minimum of the same energy level again, the two quantum states corresponding to the local minima are different. Such instances are rare and occur only when there is intrinsic degeneracy of the local minimum.

\begin{figure}
     \centering
     \begin{subfigure}{0.49\textwidth}
         \centering
         \includegraphics[width=7.5cm]{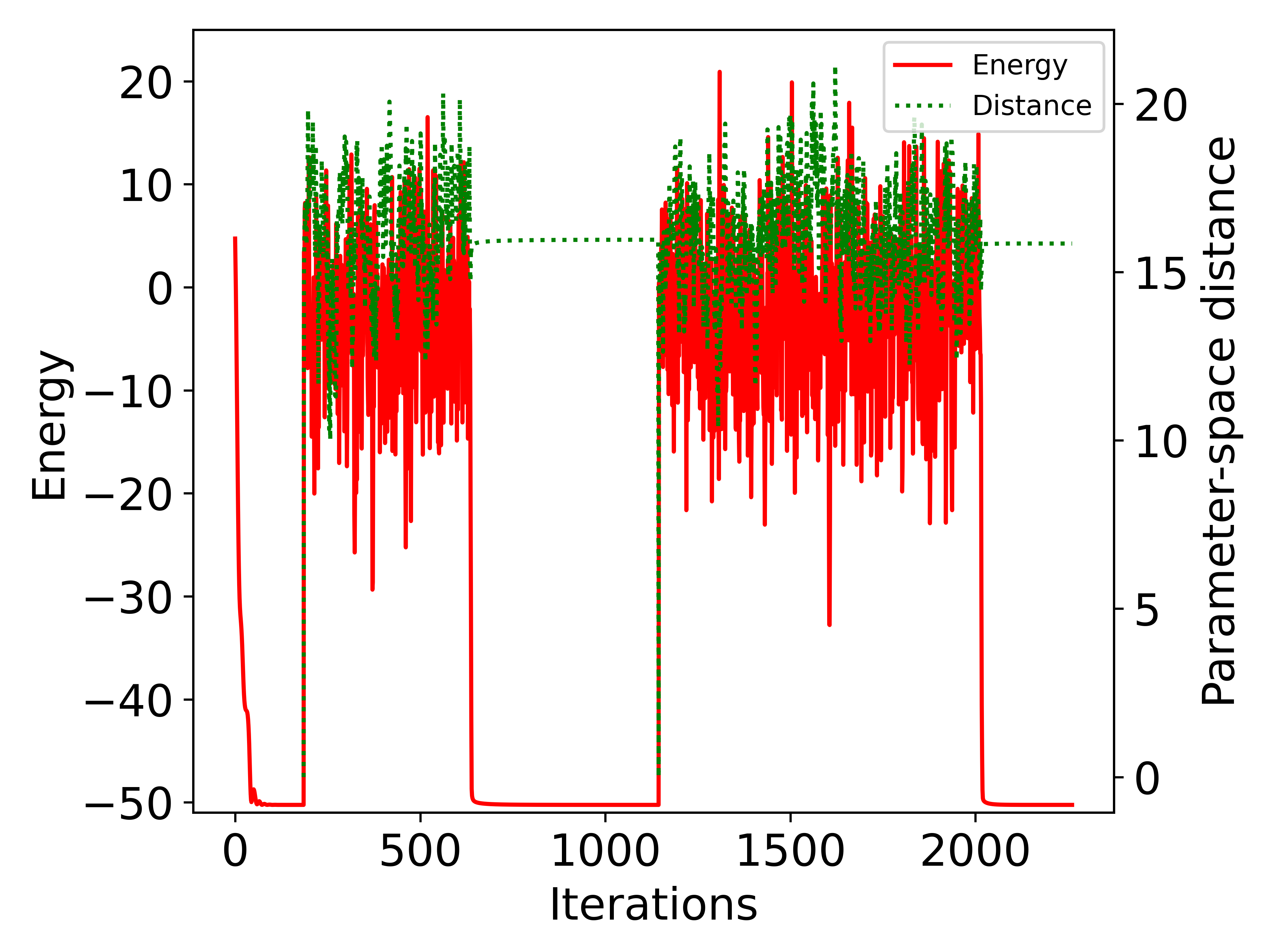}
        \caption{}
        
     \end{subfigure}
      \hfill
     \begin{subfigure}{0.49\textwidth}
         \centering
         \includegraphics[width=7.5cm]{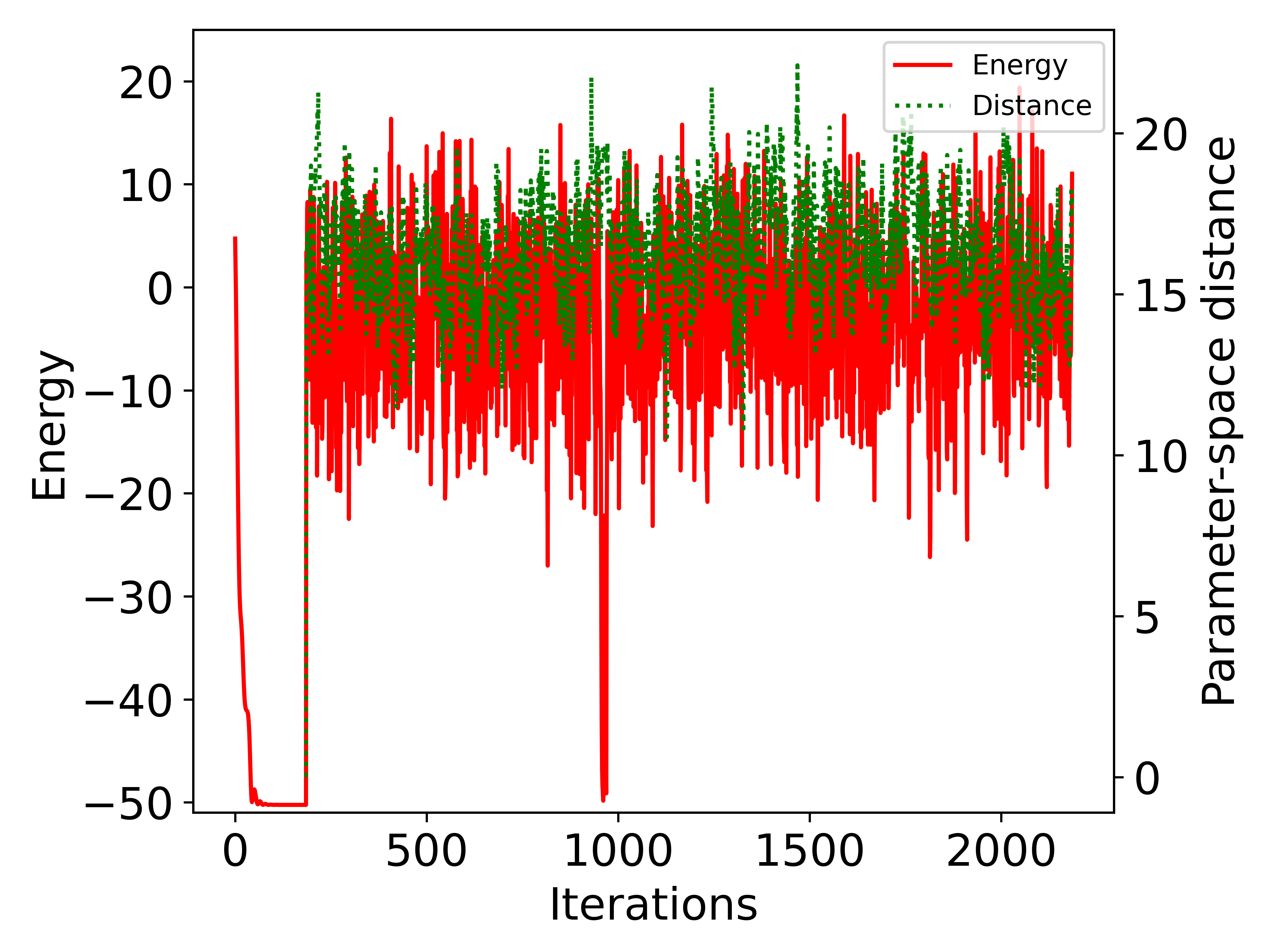}
        \caption{}
       
     \end{subfigure}
      
\caption{The iteration profiles around the \emph{global} minimum of the (a) conventional and (b) modified dynamic tunneling methods. The green dashed lines show the value of the parameter-space distance at the corresponding iteration steps, and we have put an ad-hoc bound $2^\lambda$ on the $\lambda$ term for the conventional method. The conventional dynamic tunneling flow never converges to a single global minimum, and the iteration must be terminated manually.}
\label{grsameqstates}
\end{figure}

\Fref{grsameqstates} highlights the worst-case examples that are not featured in \Fref{convplot} or \ref{lmsameqstates}.
Recall that there are multiple global minima corresponding to identical ground states but located at different points in the parameter space. As shown in \Fref{grsameqstates}(a), the flow in the conventional tunneling method may jump between these alternative global minima, never converging to a single minimum. In our simulation, this happened 23 times out of 300 samples starting from the global minimum. In these instances, the termination criterion fails to determine the global minimum, even with an increase in the iteration limit. On the other hand, when using the modified tunneling method, the flow cannot converge into other global minima, as shown in \Fref{grsameqstates}(b). This ensures the termination criterion works well, allowing for the safe determination of the global minimum. 

\section{Conclusion and outlook}
\label{sec:outlook}

In this work, we have proposed a global optimization algorithm for VQA by employing the dynamic tunneling flow generated by \eref{eq:tunneling_flow_vqe}. In contrast to the conventional dynamic flow, our dynamic flow by \eref{eq:tunneling_flow_vqe} exploits the distance measure of quantum states to resolve the extrinsic degeneracy on the quantum function arising from the parametrization of the ansatz state.

The modified dynamic tunneling method has been applied to the VQE for the transverse-field Ising model.
Our simulation results 
demonstrate the enhanced performance of the modified dynamic tunneling method as a global optimization algorithm,
in terms of the convergence to the global solution,
the analytic behavior of the method during the process, and 
the number of iterations to reach the global solution.

As indicated in \eref{eq:tunneling_flow_vqe}, the dynamic tunneling flow for VQA heavily relies on a distance measure on quantum states. Therefore, a further investigation of the distance measure concerning a practical implementation of the flow is required in addition to the direct adoption of known efficient measures computed by the quantum computation framework \cite{flammia2011, kuzmak2021measuring, lee2003}.

Finally, the success of the algorithm is also sensitive to an appropriate choice of the value of the hyperparameters. While the hyperparameter $\lambda$ in \eref{eq:tunneling_flow_vqe} can be determined from the history of gradients of the quantum function, the tunneling penalty $k$ of the flow remains to be suitably chosen to ensure that the flow converges to the lower valley. If the target system is scalable, one can devise a strategy to estimate the appropriate value of $k$ based on the dimension of the space of variational parameters. During the simulation, we extrapolate the value of $k$ by gradually increasing the system size $N$, starting from a sufficiently small size, $N=6$. 
In general, the tunneling penalty term arises from the constraint imposing the feasible space of the flow to be below the known local minima. Thus, one can regard the flow as constrained dynamics, where optimization routines for constrained problems, such as the interior-point method \cite{wachter2006implementation}, can be applied to improve the convergence of the flow instead of delicately choosing $k$.

\ack
We are grateful to Jeongho Bang for valuable comments on the draft. S.P. and M.-S.C. acknowledge the support of the Ministry of Science and ICT of Korea (Grant Nos. 2022M3H3A106307411 and 2023R1A2C1005588). Also, K.B. and S.L. acknowledge the support of the Ministry of Science and ICT of Korea by the National Research Foundation of Korea (NRF-2022M3E4A1077094, RS-2023-00281456, NRF-2024M3K5A1004355) and the Institute of Information and Communications Technology Planning and Evaluation grant funded by the Korean government (2019-0-00003, ``Research and Development of Core Technologies for Programming, Running, Implementing and Validating of Fault-Tolerant Quantum Computing System'').

\appendix


\section{Anti-ferromagnetic phase}
\label{sec:appa}

In the main text, we mainly focused on the paramagnetic phase of the transverse-field Ising model to simplify the discussion and make the main points clearer. In this appendix, for reference, we provide the simulation results in the anti-ferromagnetic phase. Exhibiting an \emph{intrinsic} ground-state degeneracy and long-range correlations, the anti-ferromagnetic phase sets a further testing ground compared to the paramagnetic phase. 

We set $N=10$ and $B=0.5$ in the model \eref{eq:tfi_1d}, and use the same ansatz as in the paramagnetic phase, as described in \Fref{ansatz}. Due to the finite size, the ground states are nearly degenerate with energies around -9.76, as explained in \Sref{sec:examples}.

\begin{figure}
     \centering

     \begin{subfigure}{0.49\textwidth} 
         \centering
         \includegraphics[width=7.5cm]{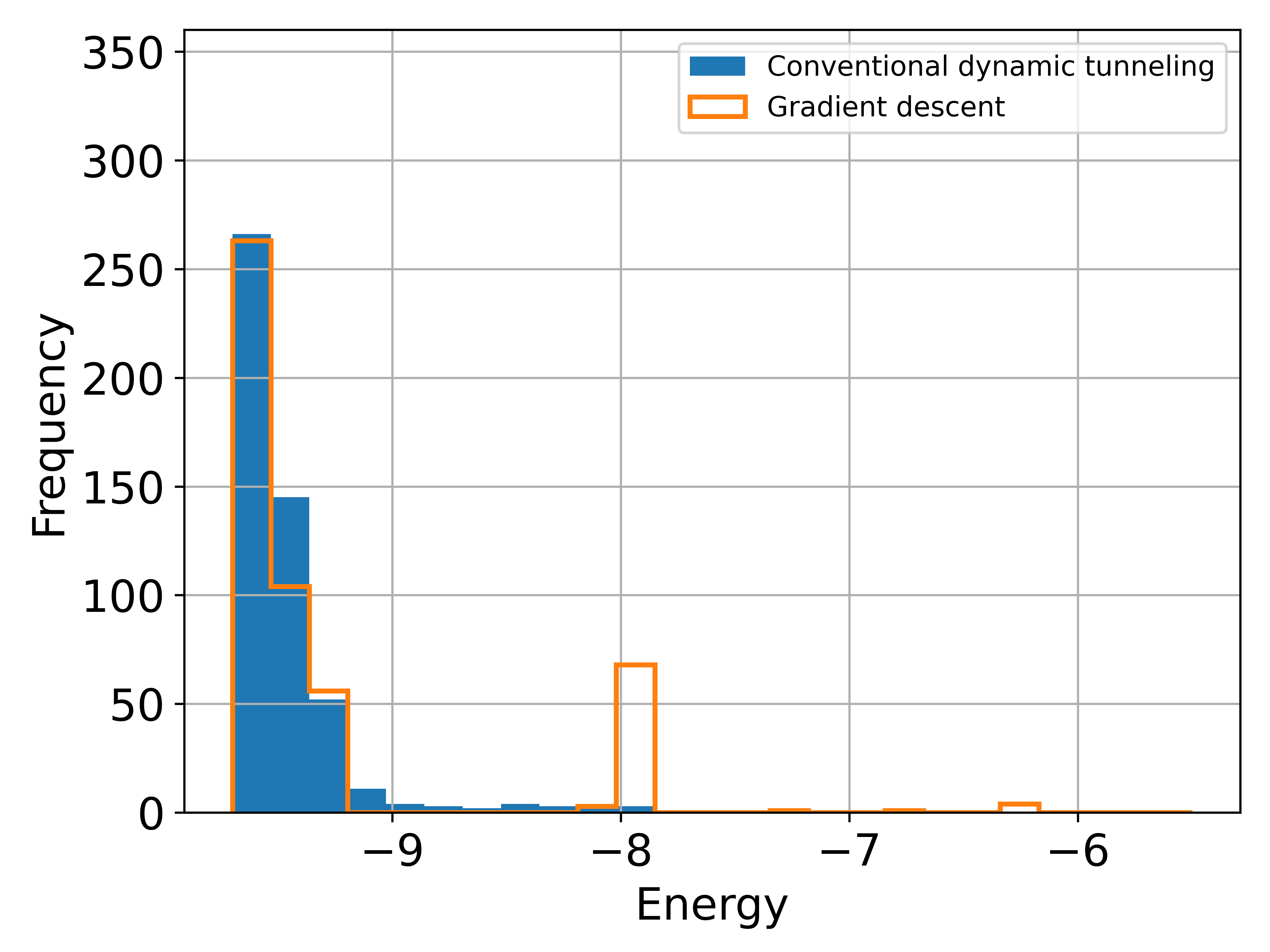}
        \caption{}
        
     \end{subfigure}
     \hfill
     \begin{subfigure}{0.49\textwidth} 
         \centering
         \includegraphics[width=7.5cm]{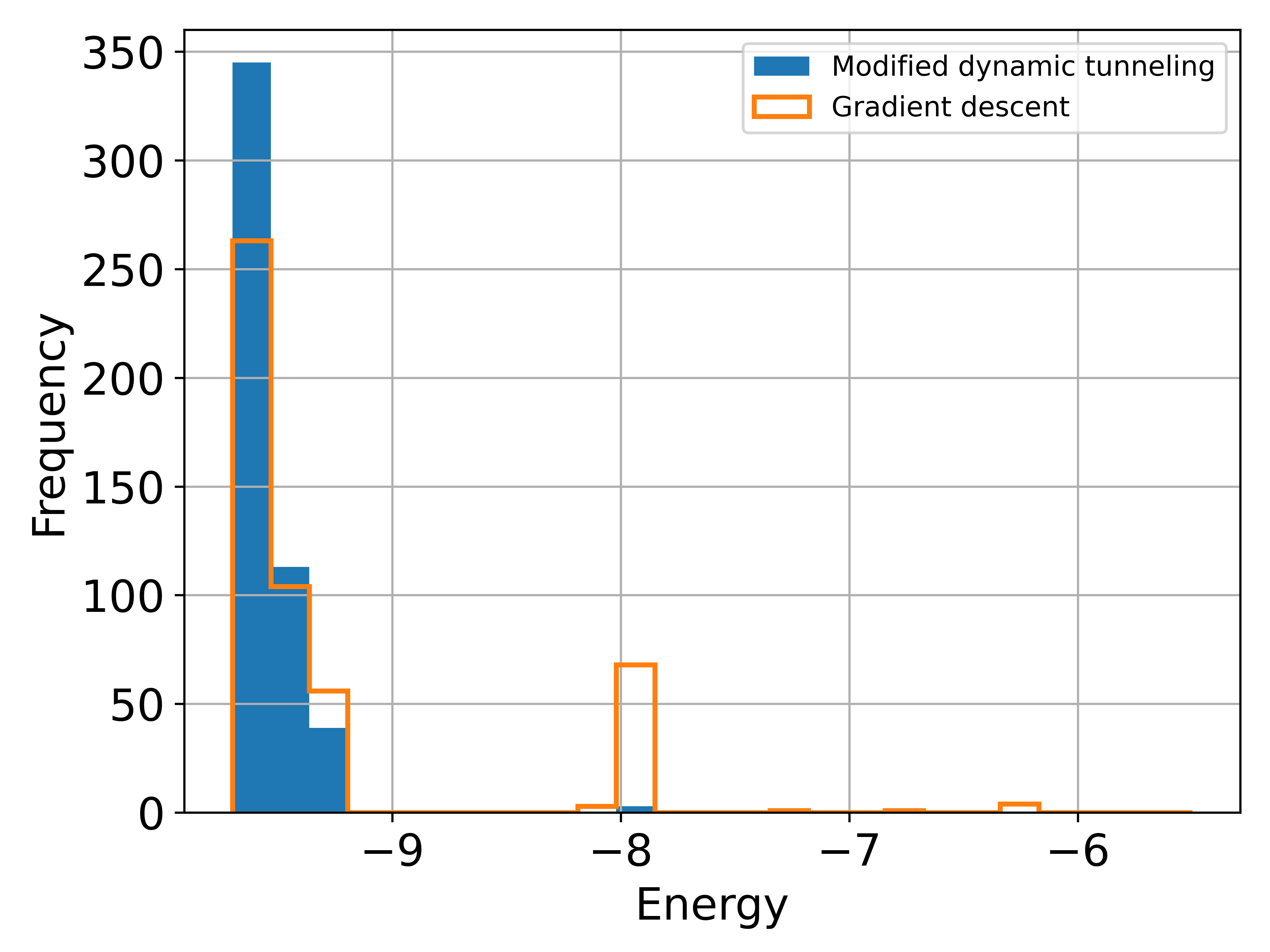}
       \caption{}
       
     \end{subfigure}
     
\caption{A comparison of the overall convergence of the (a) conventional and (b) modified dynamic tunneling method. Each histogram shows the final converged values of the cost function over 500 random samples. 
The histogram with empty bars and orange boundaries from the simple gradient-descent method is included as a reference. The global minimum is approximately -9.76 in this example.}
\label{hist_ferro}
\end{figure}

\Fref{hist_ferro} shows the overall convergence of different methods. As in the paramagnetic phase (\Fref{hist}), the modified dynamic tunneling method better escapes local minima and reaches the subspace of the nearly degenerated ground states faster. Note that the few distinct bins near the global minimum are different linear combinations of the nearly degenerate ground states, which are difficult to distinguish within the ansatz subspace. 

Technically, during the optimization process, we used FISTA\cite{beck2009fast} (a variant of gradient descent optimizer) to accelerate the update of the parameters for both the conventional and modified dynamic tunneling methods. This is necessary because in the anti-ferromagnetic phase, the gradient is significantly smaller than in the paramagnetic case across the entire cost function landscape. On the other hand, we keep the same hyperparameters except for the value of $k$, which is set to $16$ in this case.

\begin{figure}
     \centering
     \begin{subfigure}{0.49\textwidth} 
         \centering
         \includegraphics[width=7.5cm]{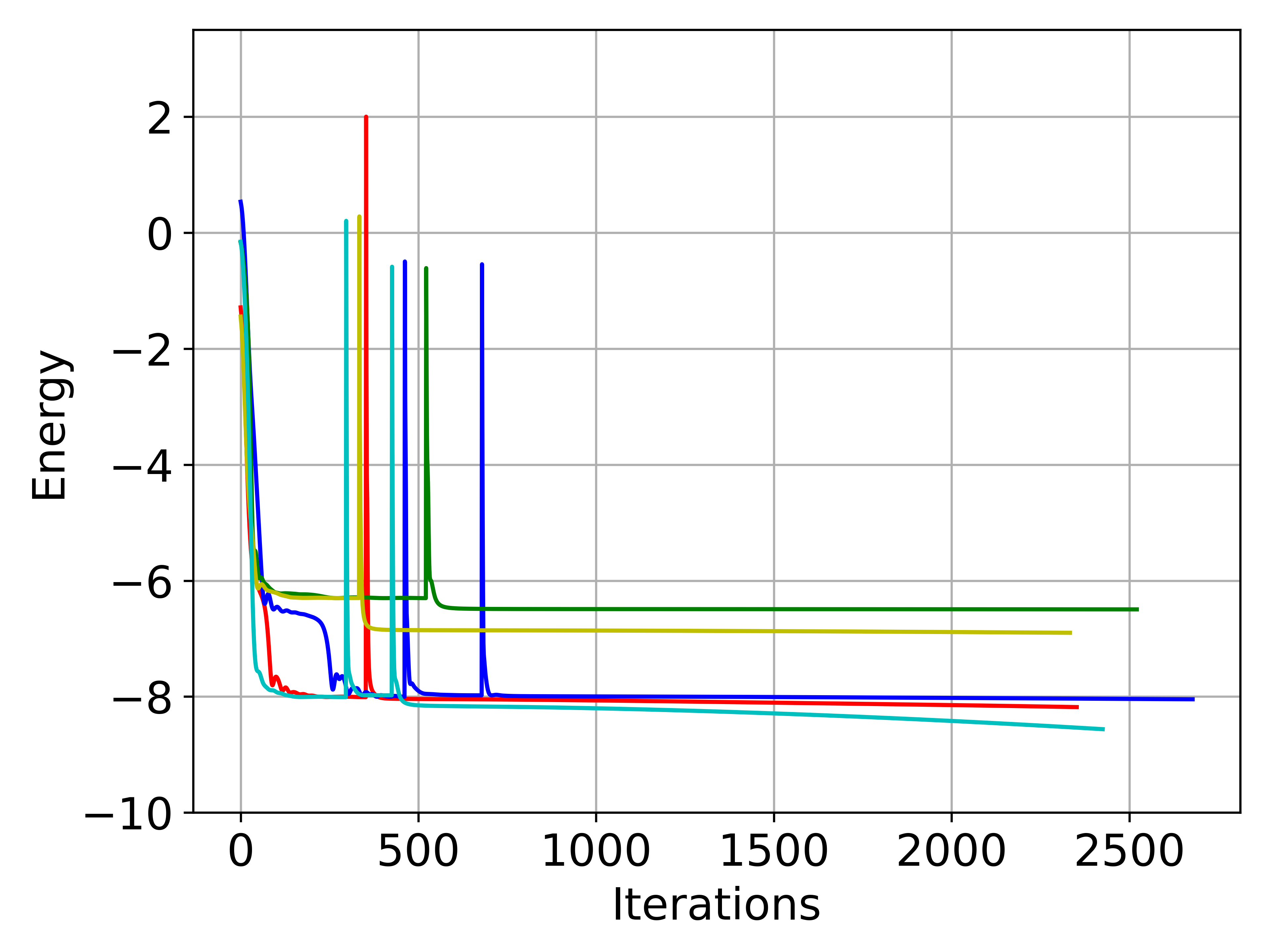}
        \caption{}
        
     \end{subfigure}
      \hfill
     \begin{subfigure}{0.49\textwidth}
         \centering
         \includegraphics[width=7.5cm]{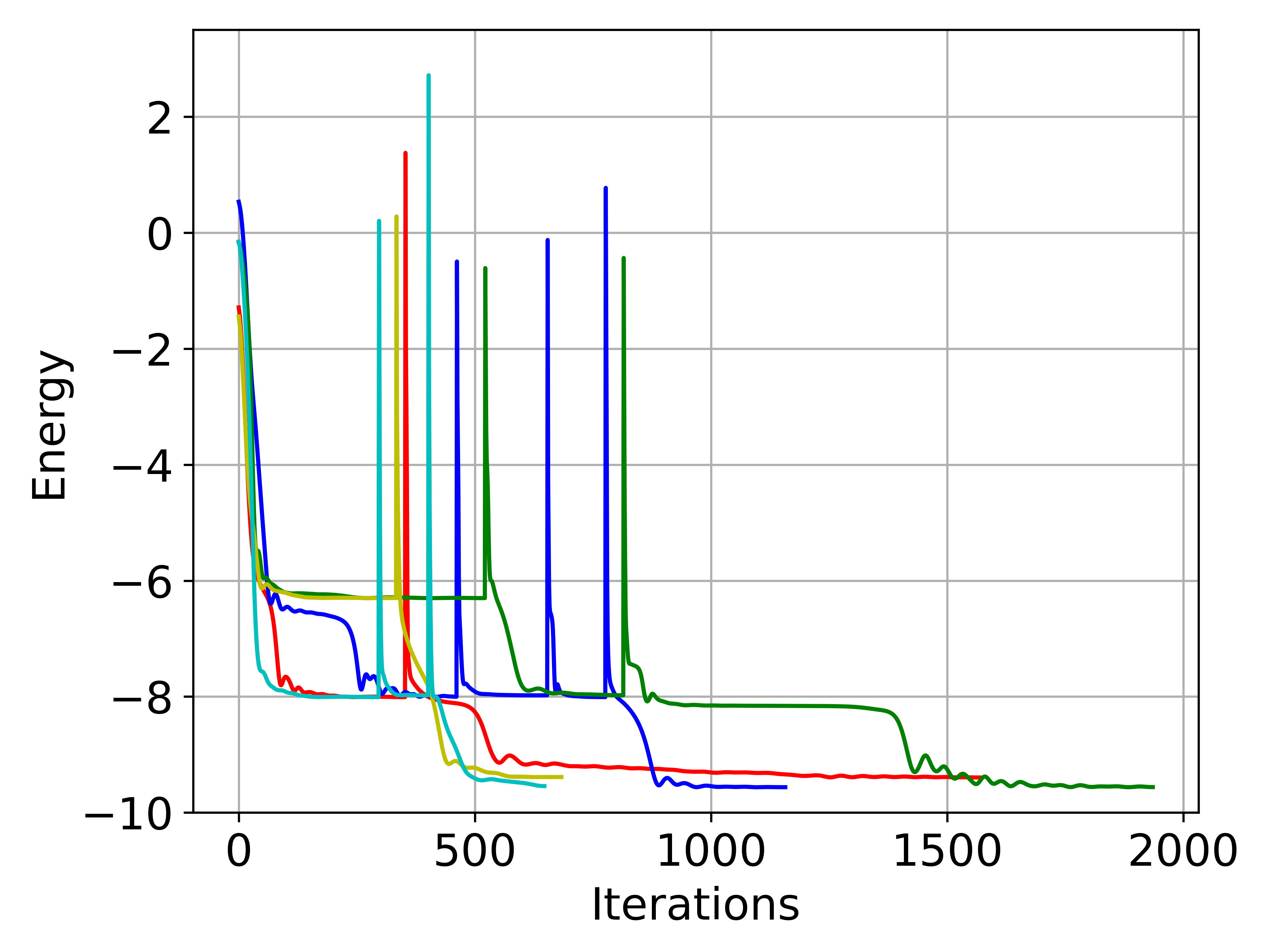}
        \caption{}
        
     \end{subfigure}
\caption{Iteration profiles of the cost function (energy) over the entire optimization process in the (a) conventional and (b) modified dynamic tunneling methods for 5 representative samples out of the whole in \Fref{hist_ferro}.}
\label{convplot_ferro}
\end{figure}

\Fref{convplot_ferro} presents the profiles of the cost function of some representative samples throughout the entire optimization process. As in the paramagnetic phase, the optimization process of the conventional dynamic tunneling method in \Fref{convplot_ferro}(a) stops on a slope, failing to converge into any minima. In contrast, as shown in \Fref{convplot_ferro}(b), the modified dynamic tunneling method successfully finds the global minimum within fewer iteration steps, despite the intrinsic degeneracy causing the iterations to encounter local minima with the same energy level multiple times.

\section{Higher-dimensional models}

In the main text, we have discussed 1D transverse-field Ising model to compare the performances of the conventional and modified dynamic tunneling methods. In this appendix, for completeness, we provide some simulation results for the transverse-field Ising model on a two-dimensional square lattice. From the results in the main text and below, we expect the same conclusion in three dimensions.

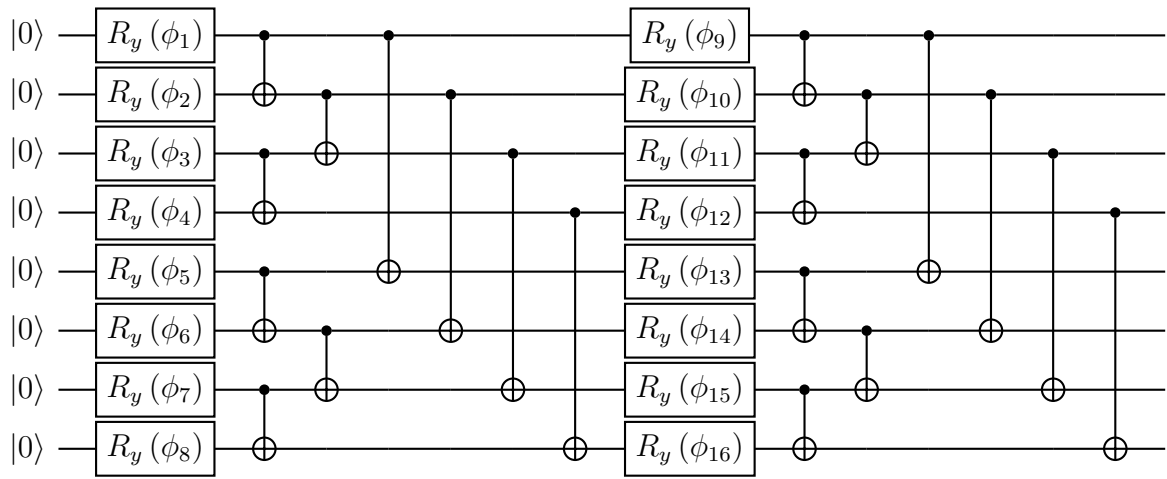
\begin{figure}

\begin{center}
    \begin{quantikz}[row sep = 2]
        \lstick{\ket{0}} & \gate{R_{y}\left( \phi_{1} \right)} & \ctrl{1} & \qw & \ctrl{4} &\qw &\qw &\qw & \gate{R_{y}\left( \phi_{9} \right)} & \ctrl{1} & \qw & \ctrl{4} &\qw &\qw &\qw  &\qw \\
        \lstick{\ket{0}} & \gate{R_{y}\left( \phi_{2} \right)} & \targ{ } & \ctrl{1} &\qw &\ctrl{4} &\qw &\qw  & \gate{R_{y}\left( \phi_{10} \right)} & \targ{ } & \ctrl{1} &\qw &\ctrl{4} &\qw &\qw&\qw\\
        \lstick{\ket{0}} & \gate{R_{y}\left( \phi_{3} \right)} & \ctrl{1} & \targ{ } &\qw&\qw&\ctrl{4}&\qw & \gate{R_{y}\left( \phi_{11} \right)} & \ctrl{1} & \targ{ } &\qw&\qw&\ctrl{4}&\qw&\qw\\
        \lstick{\ket{0}} & \gate{R_{y}\left( \phi_{4} \right)} & \targ{ } & \qw &\qw&\qw&\qw&\ctrl{4}  & \gate{R_{y}\left( \phi_{12} \right)} & \targ{ } & \qw &\qw&\qw&\qw&\ctrl{4}&\qw\\
        \lstick{\ket{0}} & \gate{R_{y}\left( \phi_{5} \right)} & \ctrl{1} &\qw  &\targ{} &\qw &\qw &\qw   &\gate{R_{y}\left( \phi_{13} \right)} & \ctrl{1} &\qw  &\targ{} &\qw &\qw &\qw&\qw   \\  
        \lstick{\ket{0}} & \gate{R_{y}\left( \phi_{6} \right)} & \targ{ } &\ctrl{1} &\qw &\targ{}  &\qw &\qw   &\gate{R_{y}\left( \phi_{14} \right)} & \targ{ } &\ctrl{1} &\qw &\targ{}  &\qw &\qw&\qw    \\ 
        \lstick{\ket{0}} & \gate{R_{y}\left( \phi_{7} \right)} & \ctrl{1} & \targ{ }&\qw&\qw&\targ{} &\qw   & \gate{R_{y}\left( \phi_{15} \right)} & \ctrl{1}      & \targ{ }&\qw&\qw&\targ{} &\qw&\qw \\
        \lstick{\ket{0}} & \gate{R_{y}\left( \phi_{8} \right)} & \targ{ } & \qw&\qw&\qw&\qw&\targ{}  & \gate{R_{y}\left( \phi_{16} \right)}      & \targ{ } & \qw&\qw&\qw&\qw&\targ{} &\qw
    \end{quantikz}
\end{center}

\caption{A parameterized quantum circuit corresponding to a variational ansatz for the transverse-field Ising model on a $2\times 4$ ($N=8$) lattice. The ansatz may be constructed in a similar fashion for larger lattices. $R_y(\phi_i):=e^{-i \sigma^y \phi_i/2}$ denotes the single-qubit rotation around the $y$-axis.}
\label{ansatz_2d}
\end{figure}

We consider a $3\times 4$ square lattice ($N=12$) and set $B=5$. The ansatz is slightly modified to the form depicted in \Fref{ansatz_2d} [on a $2\times 4$ lattice ($N=8$) for illustration] in order to account for the connectivity between qubits on the lattice.

The hyperparameters for this simulation are set to be the same as those for the 1D cases, except for the value of $k$, which is set to $0.35$ now. We again use the accelerated gradient-descent method to enhance the overall convergence rate for both the conventional and modified dynamic tunneling methods.

\begin{figure}
     \centering

     \begin{subfigure}{0.49\textwidth} 
         \centering
         \includegraphics[width=7.5cm]{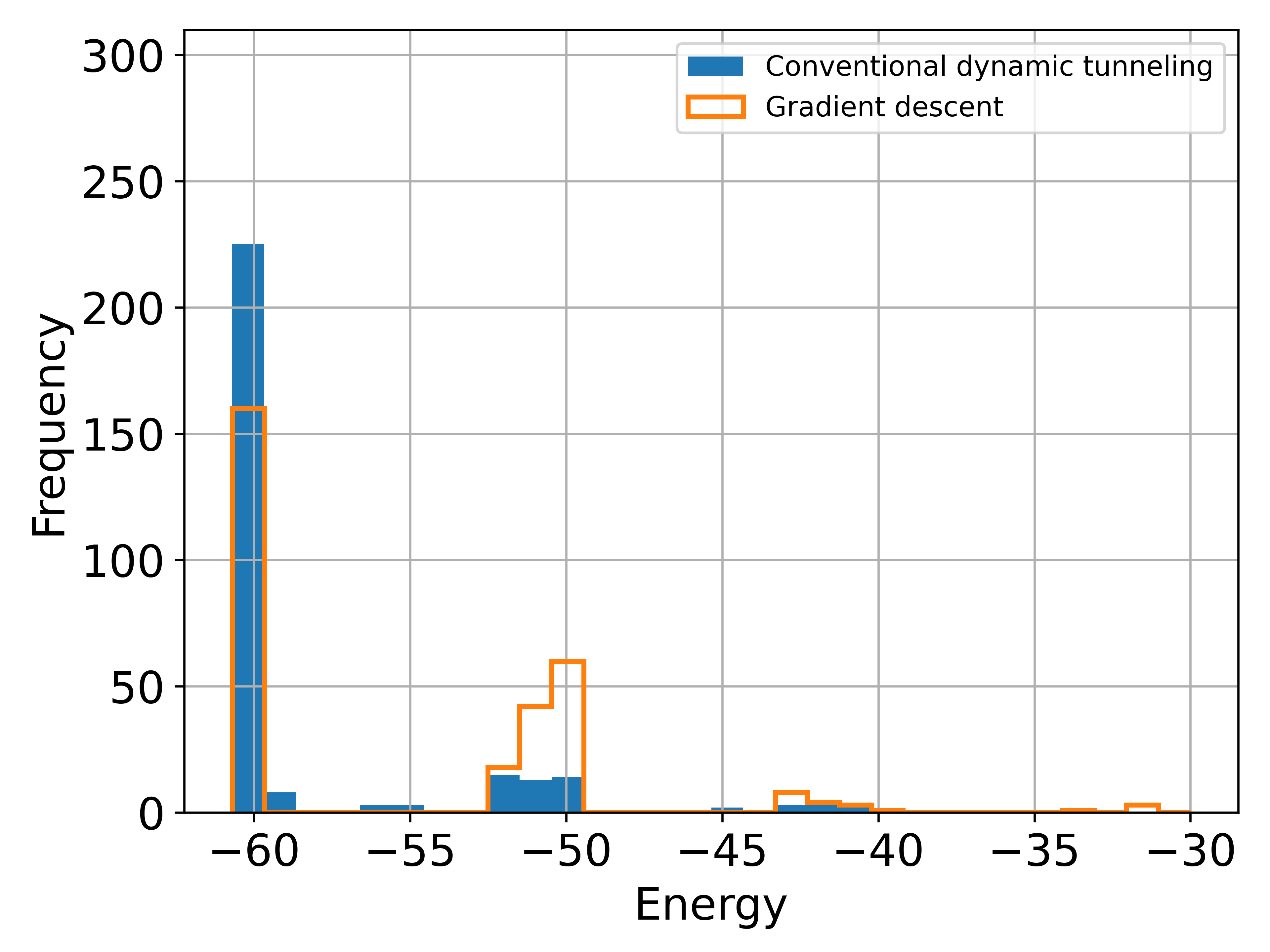}
        \caption{}
        
     \end{subfigure}
     \hfill
     \begin{subfigure}{0.49\textwidth} 
         \centering
         \includegraphics[width=7.5cm]{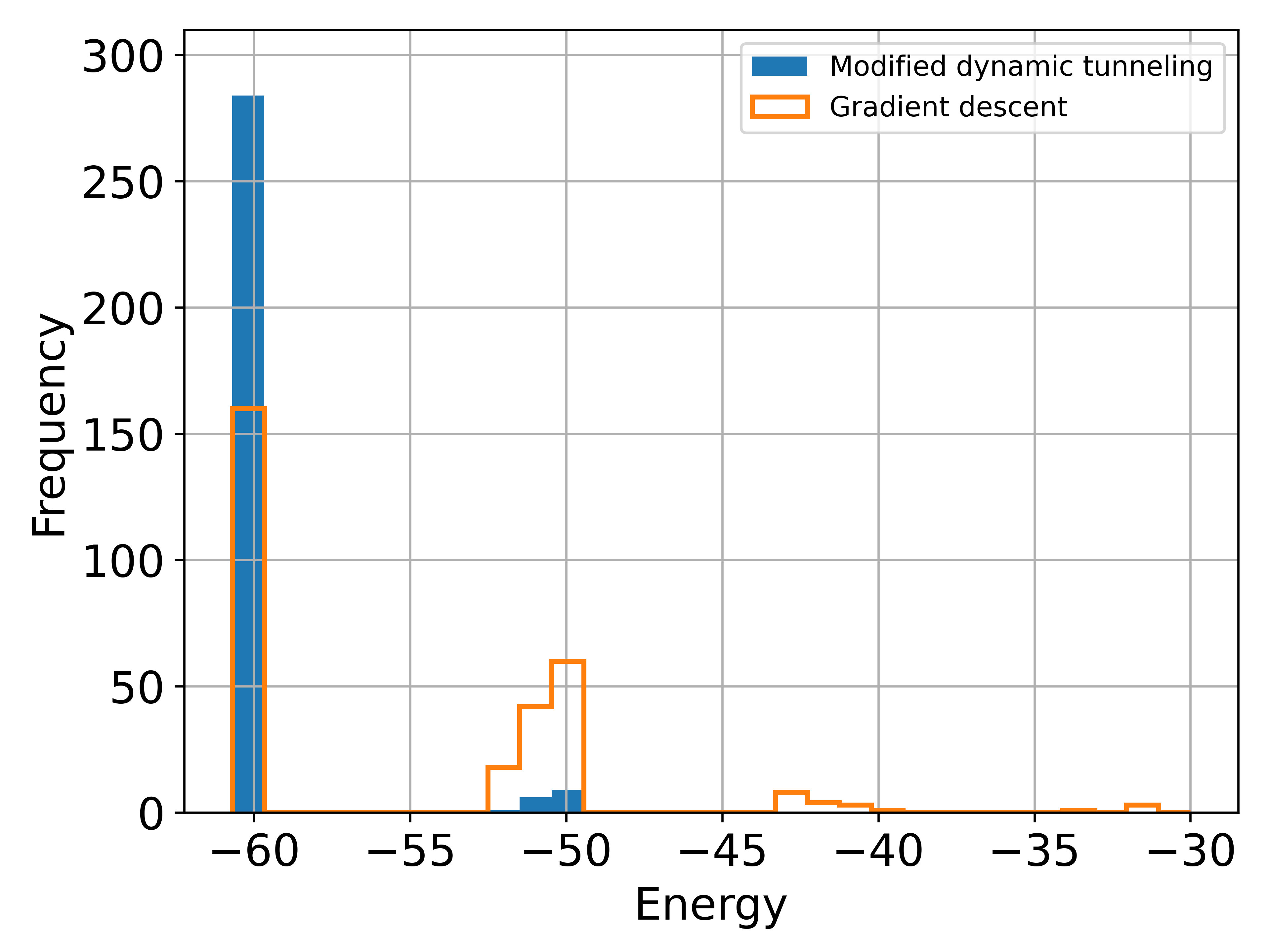}
       \caption{}
       
     \end{subfigure}
     
\caption{Overall convergence of the (a) conventional and (b) modified dynamic tunneling methods. Each histogram shows the final converged values of the cost function over 300 random samples.
For a reference, we also show the histogram with empty bars and orange boundaries from the simple gradient-descent method. The global minimum is approximately -60.87 in this example.}
\label{hist_2d}
\end{figure}

\Fref{hist_2d} compares the performance of our method with the conventional dynamic tunneling method. The modified dynamic tunneling method better escapes local minima and converges faster to the global minimum than the conventional one. Through this example, we expect that our method can be applied to higher-dimensional problems, which have an even more complicated landscape of the cost function.

\section*{References}

\bibliographystyle{iopart-num}
\bibliography{dyopt}

\end{document}